\documentstyle[12pt,aaspp4]{article}

\def\insertplot#1#2#3#4#5#6#7{
\vskip 10pt\nobreak\hbox to \hsize{\hss\dimen0=#3in\hbox to #6\dimen0{%
\dimen0=#2in\vbox to #6\dimen0{\vss
\special{ps: plotfile #1}
\special{ps::[end]
  PGPLOT restore
}
}\hss}\hss}\vskip 10pt}

\tighten
\lefthead{Natta, Meyer \& Beckwith}
\righthead{Silicates in T Tauri stars}

\newcommand{\simless}{\raisebox{-.8ex}{$\buildrel{\textstyle<}\over\sim$}}

\newcommand{\simgreat}{\raisebox{-.8ex}{$\buildrel{\textstyle>}\over\sim$}}
\def\Lk {LkH$\alpha$~332-20}
\def\Gl {Glass~Ia}
\def\CT {CT~Cha}
\def\XX {XX~Cha}
\def\VZ {VZ~Cha}
\def\WW {WW~Cha}
\def\SX{SX~Cha}
\def\WX {WX~Cha}
\def\VW {VW~Cha}

\def\pms {pre-main--sequence}

\def \Lstar {L$_\star$\,}
\def \Rstar {R$_\star$\,}
\def \Tstar {T$_\star$\,}
\def \Mstar {M$_\star$\,}

\def\Lsun {L$_\odot$\,}
\def\Msun {M$_\odot$\,}

\def\um {$\mu$m}

\begin{document}

\title{Silicate Emission in T Tauri Stars:  Evidence for Disk Atmospheres? \altaffilmark{1}}

\author{Antonella Natta}
\affil{Osservatorio Astrofisico di Arcetri, Largo Enrico
Fermi 5, I--50125 Firenze, Italy}

\author{Michael R. Meyer\altaffilmark{2}}
\affil{Steward Observatory, The University of Arizona, 933 N. Cherry Ave., \\ 
Tucson, AZ 85721--0065}

\and

\author{Steven V.W. Beckwith\altaffilmark{3}}
\affil{Space Telescope Science Institute, 3700 San Martin Drive,
Baltimore, MD  21218}

\altaffiltext{1}{ Based on observations obtained with ISO. ISO is an ESA project
with instruments funded by ESA Member States (especially the PI
countries: France, Germany, the Netherlands and the United Kingdom)
and with the participation of ISAS and NASA.}

\altaffiltext{2}{Hubble Fellow}

\altaffiltext{3}{also at the Max--Planck--Institut f\"ur Astronomie, K\"onigstuhl 17, Heidelberg,
Germany, D--69117}

\begin{abstract}

We present low-resolution mid--infrared spectra of nine classical T Tauri stars
associated with the Chamaeleon I dark cloud.  The data were  obtained with
the PHOT-S instrument on--board the Infrared Space Observatory (ISO)
in the two wavelength ranges 2.5--4.9 and 5.9--11.7 \um.
All nine stars show evidence of silicate emission
at 10 \um, which is the only prominent feature
in the spectra. We discuss a model for the origin of these features in a hot
optically--thin surface layer of the circumstellar disks surrounding
the central young stars (i.e. a disk atmosphere).
We report excellent agreement of our observations with
predictions based upon this simple model for most
stars in our sample, assuming that a mixture of amorphous silicates
of  radius $\simless 1$ \um\
is the dominant source of opacity.  These observations
support the notion that extended disk atmospheres contribute
substantially to the mid--IR flux of young stars.

\end{abstract}


\section{INTRODUCTION}

It is generally recognized that circumstellar disks provide the most
economical means of accounting for the spectral energy distributions
of T Tauri stars.  The extended, flat distributions of gas and dust
can easily generate the observed broad spectral energy distributions
(Lynden-Bell \& Pringle 1974; Adams, Lada, \& Shu 1987), and
calculations of disk spectra have been successful at reproducing
many observations of these stars (Calvet et al. 1992; D'Alessio et al. 1999;
Chiang \& Goldreich 1997, 1999).  In some cases, the disks are apparent in
images of the stars, either via scattered light or in silhouette against a
background light source (Burrows et al. 1996; McCaughrean \& O'Dell 1996).

The disks may be heated either radiatively by the central star
or by energy released as matter accretes through the disk driven
by turbulent viscosity.   It is still a matter of debate which of
these is more important for any given star, and
the variety of spectral energy distributions observed indicates that the
principal heating source may be different in various star+disk systems.  It
is important to determine the extent to which radiation or
accretion dominates, because almost all other aspects of our understanding
of these disks depends on the energy budget as a function of radius.
For example, the amount of convection will depend on the
vertical temperature gradient in the disk atmosphere.

A clear distinction between an atmosphere heated from the top by radiation
and from the bottom by accretion is the presence of spectral features.
An atmosphere heated from above by radiation will generally produce
emission features, unless the viewing angle is very close to the plane
of the disk.  A disk heated from the mid--plane by accretion will have an
atmosphere with a vertically decreasing temperature, and all spectral
features will be in absorption. Observations of disk spectra should
provide a good discriminant between these two possibilities.
For example, FU Ori objects, a class of eruptive variables whose
luminosities are dominated by accretion,  exhibit deep absorption
features due to their low surface gravity disk atmospheres
(Kenyon \& Hartmann, 1997).

Cohen \& Witteborn (1985) showed that some T Tauri stars had broad
resonance features near 10 $\mu$m, although it was not known
at the time where these lines were formed and the signal--to--noise
ratios were not high.  We set out to verify if these emission features are
the rule by observing a set of T Tauri stars with the spectrometer
of ISOPHOT, PHOT-S, on the Infrared Space Observatory.  In section II,
we describe the observations and analysis of the data, while in
section III we review the results.  In section IV, we discuss the nature
of the observed silicate feature with emphasis on comparison to predictions
of the disk atmosphere model.
Finally, in section V we summarize our results.

\section {OBSERVATIONS AND DATA REDUCTION}

We have selected a sample of nine well--studied TTS in the Chamaeleon I
dark cloud.  These objects are listed in Table~1 along with
relevant stellar parameters taken or derived from the published
literature.
All the stars in this sample have an infrared excess, indicating
the presence of a circumstellar disk or envelope.
Source positions were taken from Gauvin
\& Strom (1992) and checked with Lopez \& Girard (1990).
Discrepancies in position were $< 10$'' for all sources.

We obtained spectrophotometric data of our sample
during orbits 166 and 270 (on May 1 and August 14, 1996 respectively).
We utilized the PHOT--S module of the ISOPHOT instrument (Lemke et al., 1996)
providing a resolving power of $R \sim 50$ from 2.5--4.9 $\mu$m
and 5.9--11.7 $\mu$m
The observations were performed in triangular--chop mode within approved
observing template PHT40.  Sky--positions were selected 90'' north and south
of the targets and the input aperture was 28'' $\times$ 28''.  On--source
integration times were 512 seconds per source.
Dark current observations were obtained
in order to monitor detector history effects.

\begin{figure}
\insertplot{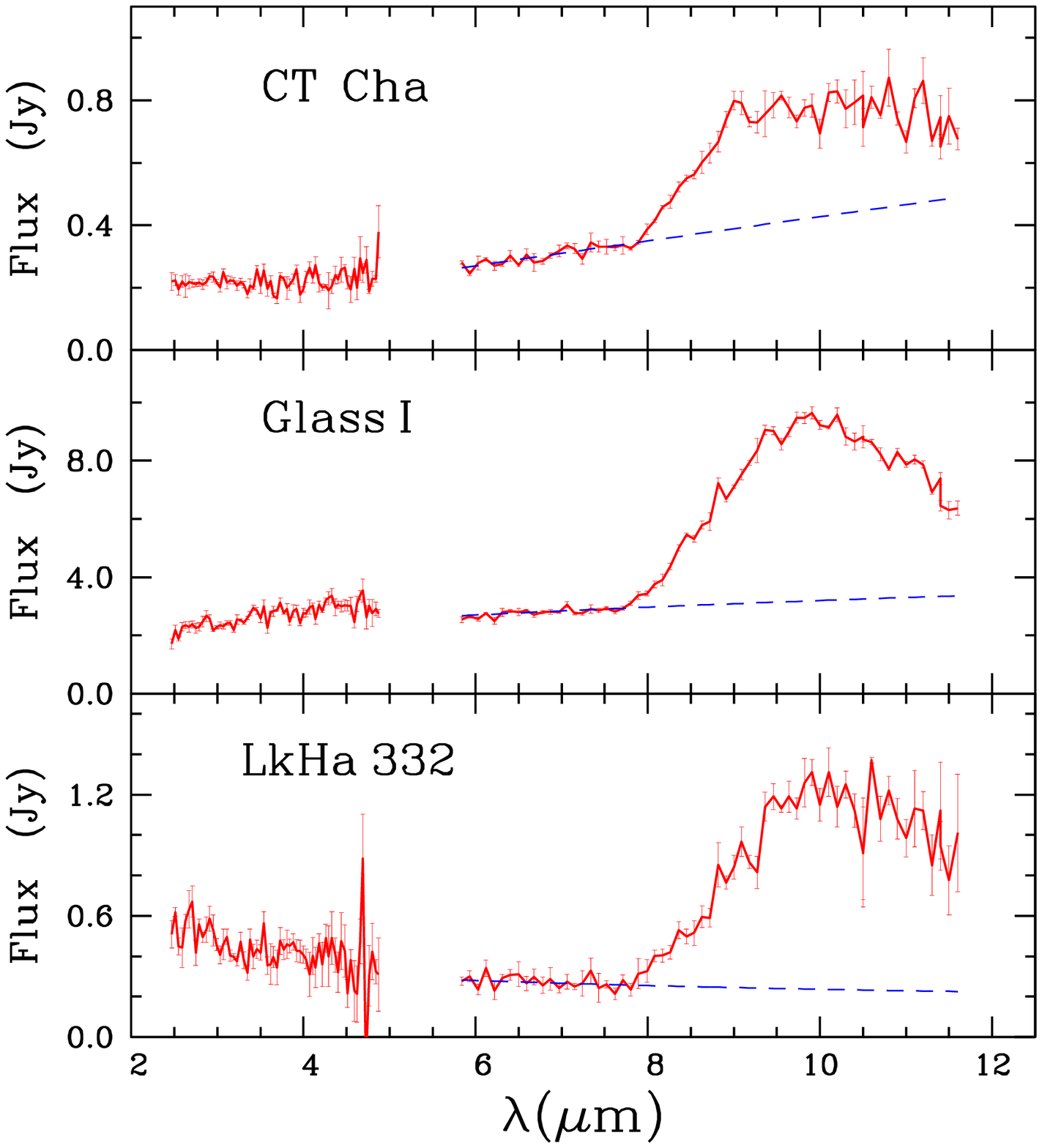}{8.0}{10.}{-0.75}{1.0}{0.8}{0}
\caption
{Observed spectra for \CT, \Gl, and \Lk.
The dashed lines in Fig.~\ref{spectra_1}--\ref{spectra_3}
show  for each star linear fits to the spectral region 5.8--8 \um.
}
\label{spectra_1}
\end{figure}

\begin{figure}
\insertplot{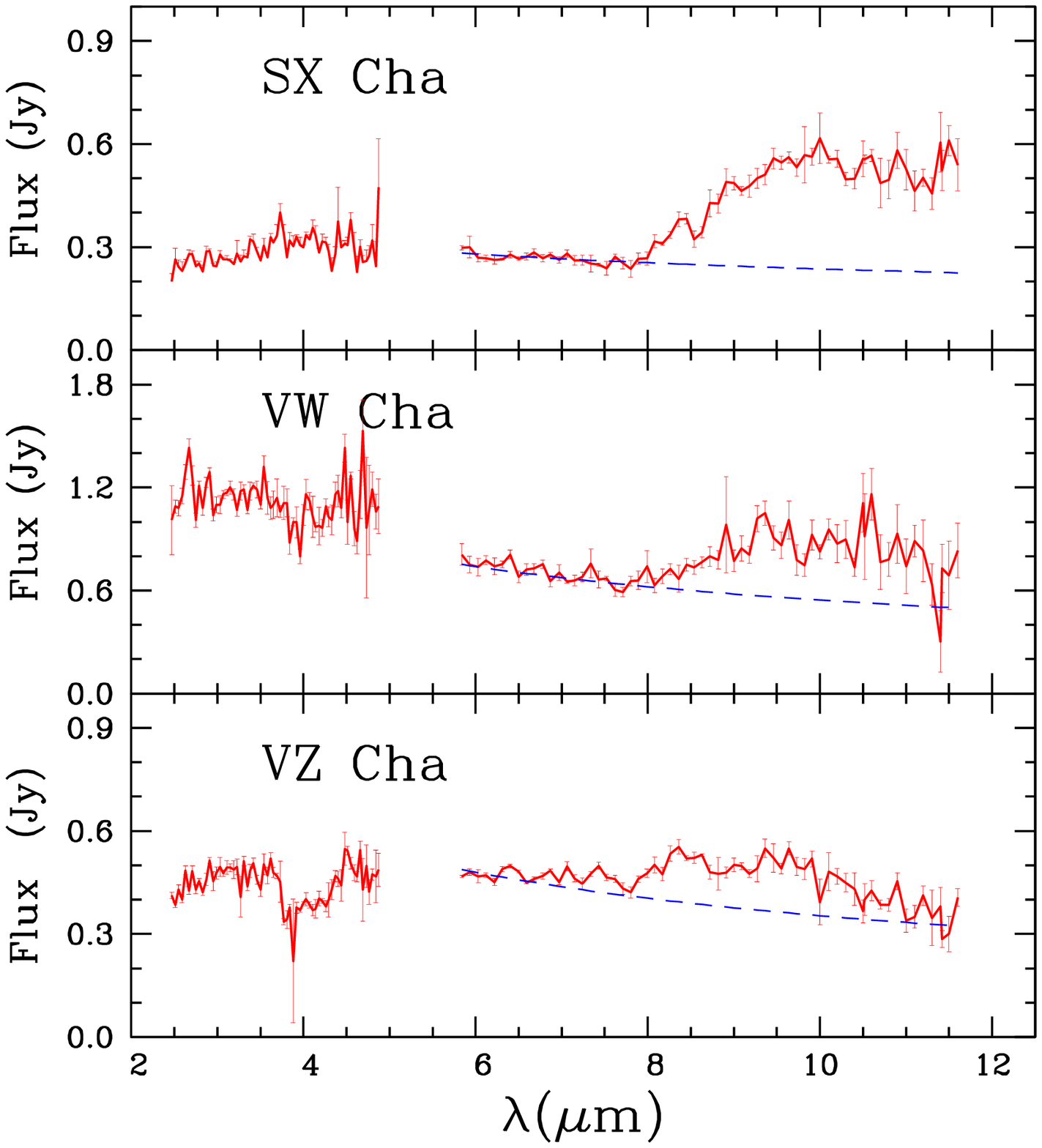}{8.0}{10.}{-0.75}{1.0}{0.8}{0}
\caption
{Same as Fig.~\ref{spectra_1} for \SX, \VW, and \VZ.}
\label{spectra_2}
\end{figure}

\begin{figure}
\insertplot{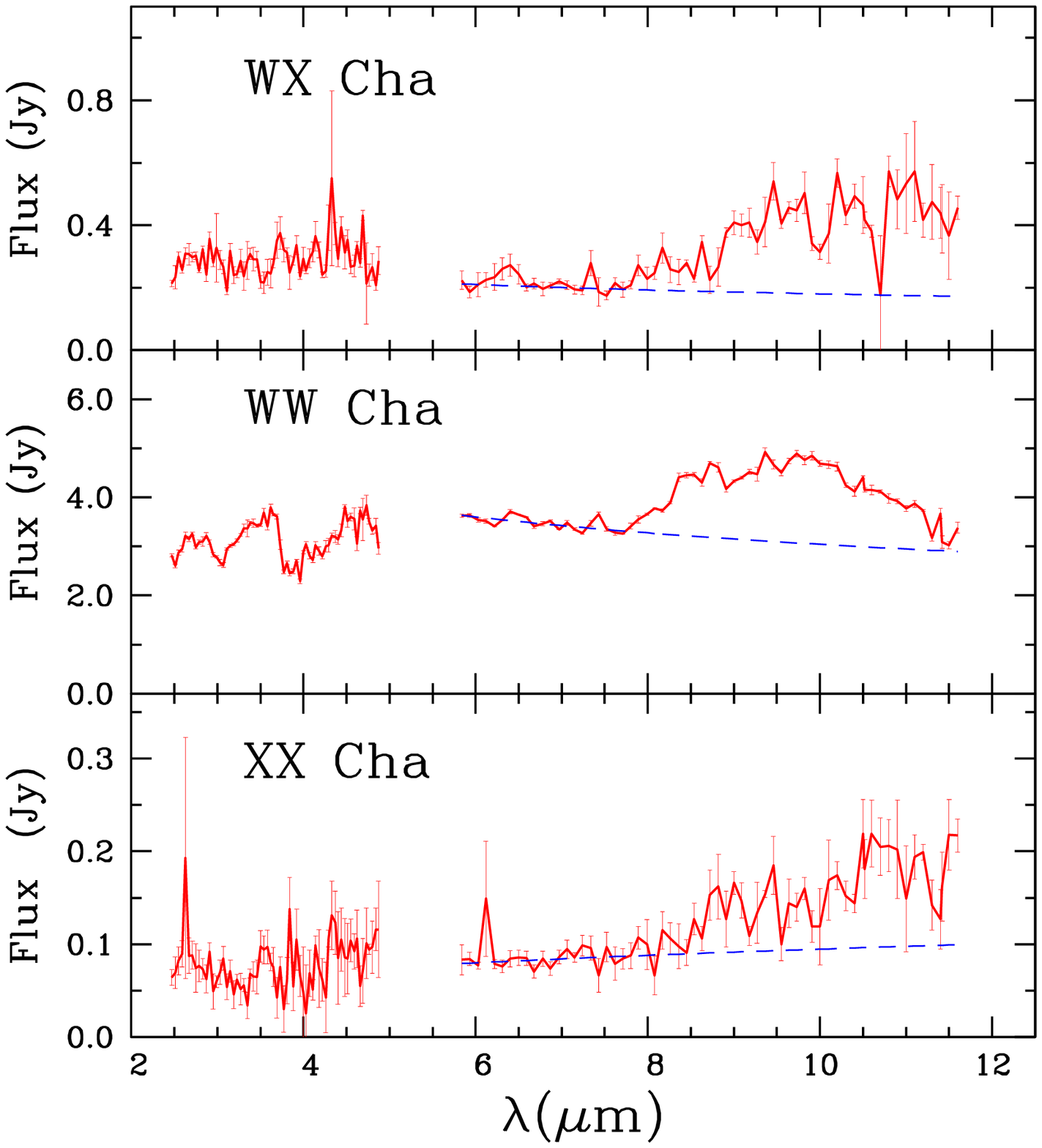}{8.0}{10.}{-0.75}{1.0}{0.8}{0}
\caption
{Same as Fig.~\ref{spectra_1} for \WX, \WW, and \XX.}
\label{spectra_3}
\end{figure}

The data were reduced using the Phot--Interactive Analysis
(PIA) Version 7.3.2 (Gabriel et al. 1997)
\footnote{PIA is a joint development by the ESA Astrophysics Division and the
ISOPHOT Consortium led by the Max--Planck--Institute for Astronomy (MPIA),
Heidelberg.}.  We utilized the two--threshold deglitching algorithm for
the edited raw data and split each signal ramp into three sub--ramps.
Slopes (in units of volts per second) were fit to each
sub--division with a linear least--squares method.  The distribution of slopes
for each measurement was checked again and spikes were discarded.
Dark subtraction was performed appropriate for the orbital time of the
observations.  The selected raw data were then averaged without attempting
to correct for the drift behavior of the detector responsivity.
Background signals were subtracted by interpolating between the
two off--source chopper positions.  The signal levels were then calibrated
using a flux--dependent grating response function derived without
drift--corrections.  The resulting spectra for the nine stars
observed are shown in Fig.~\ref{spectra_1} -- \ref{spectra_3}
plotted in units of flux density (Jy).
This calibration of the PHOT--S data available within PIA Version
7.3.2 should be accurate to $<$ 30 \% in a relative sense
and $< 30$ \% absolute (Klaas et al., 1998).
 The error bars in Fig.~\ref{spectra_1}--\ref{spectra_3}
reflect the random errors in
signal derivation and not the systematic errors in calibration.

In order to verify our results,
we reduced observations of two standard stars of different brightness;
the M0 III star HR 6464 [$F_{\nu}(10 \mu m) \sim 10.0$ Jy] and the K1 V star
HD 132142 [$F_{\nu}(10 \mu m) \sim 0.1$ Jy].  The data were processed
identically to those of our program stars and results were compared to the model
atmosphere calculations described by Cohen et al. (1996).
Comparison of the data to the predictions agreed within the errors
of the models for the bright source ($<$ 3.3  \%
per pixel) and within the errors of the observations for
the faint source ($<$ 25 \% per
pixel) over most of the 5.9--11.7 $\mu$m spectral regime.
In the bright standard, the model exhibits disagreement with the
observed spectrum near 8.3 $\mu$m where there
is a hint of an unidentified feature at a maximum depth
of 10 \%.  In the 2.5--4.9 $\mu$m band
the bright source model disagrees systematically with the
observations at $\lambda < 2.9 \ \mu$m
and between 3.9--4.2 $\mu$m ($<$ 5 \%)
near the CO fundamental bandhead at 4.6 $\mu$m.
In the faint standard, the data are
systematically lower than the model between 3.0--3.6 $\mu$m
at the level of 10 \%.
In general, these tests verify the applicability of our
signal derivation algorithm and the internal consistency of the calibration
procedure.   However, features which appear at wavelengths where the
models are clearly discrepant with the observations of the
standard stars should be treated with
suspicion.   Details concerning the calibration of PHOT--S can be found in
Acosta-Pulido et al. (2000).

\section {RESULTS}

\subsection {The short-wavelength spectrum}

The region between 2.5--4.9 \um\ is featureless (within the noise)
for most of the stars in our sample, with two possible
exceptions. In  \WW\ and \VZ\ there is some evidence of a broad
absorption feature near 4 \um.  Similar features have been observed
in a survey of K giant stars by Decin \& Vandenbussche (private communication).
Preliminary analysis suggests that they are instrumental artifacts though
this has yet to be demonstrated conclusively.  We do not see evidence for
the 3.1 $\mu$m H$_2$O nor the 4.3 $\mu$m CO$_2$ ice--band features
as seen in absorption toward heavily embedded young stellar objects (e.g.
van Dishoeck et al. 1998).

\subsection {The long-wavelength spectrum}

The region between 5.9--11.7 \um\ is characterized by the presence of the
10 \um\ silicate feature, which is seen
in emission in
most of the stars in our sample.
In addition, there is an underlying continuum emission, whose
strength (relative to the silicate feature and the stellar photosphere)
varies from star to star. We approximate this continuum
as a power-law function of wavelength normalized in
the spectral region 5.8--8 \um\ (dashed lines in
Fig.~\ref{spectra_1}--\ref{spectra_3}).

The shape of the feature varies somewhat from star to star. Three
of the best measured spectra (\Gl, \Lk\ and \WW) are quite similar,
with a peak at about 10 \um\ and a slow rise on the blue side.
The FWHM of these features after continuum subtraction is
$\sim 2.8$ \um.
Two other stars (\CT\ and \SX) have flatter
spectra with no clear peak.
We summarize the properties of the features in Table~2
which gives the name of the star, the luminosity of the
feature between 7.5 and 11.6 \um\ after subtraction of the
continuum component, the ratio of the luminosity
in the feature to the stellar luminosity, and the ratio of the
flux at the peak of the feature ($\sim$10 \um) to the continuum at the same
wavelength.
The ratio L$_{sil}$/L$_\star$ is 1--2 \%
for most stars, with the remarkable
exception of \Gl\, where it is $\sim$10\%
\footnote{However \Gl\ has an embedded companion of unknown luminosity
that may excite the silicate emission as discussed in section 4.3 below}.

G\"urtler et al. (1999) have also obtained spectra
for two of the stars in our sample, \Gl\ and \Lk.  Our observed
flux for \Gl\ is intermediate between two observations taken at
different epochs reported
in G\"urtler et al. (but closer to the one obtained nearest in time).
The observed flux for \Lk\ is similar to
that reported by G\"urtler et al. though the line shape is noticeably
different.  However,  differences in line shape could be due to
differences in reduction techniques \footnote{We employed a more recent
version of the PIA with an updated grating response function calibration.}.
There is no evidence for emission from the unidentified infrared
bands at 6.2 and 7.7 $\mu$m often attributed to polycyclic
aromatic hydrocarbons (PAHs).   Such features have been observed in
the spectra of objects that illuminate reflection nebulosities
(e.g. G\"urtler et al., 1999).
The 6.2 and 7.6 \um\ features  lie in a
region of the spectrum well measured in our data.

\section{NATURE OF THE SILICATE EMISSION}

The main interest of the silicate feature seen in these TTS spectra
lies  in the possibility it offers of investigating the
optically thin dust component of the circumstellar environment.
Historically, the feature has been described as  optically thin
emission from dust  having optical depth $\tau_\nu$ and
temperature T$_{BB}$ (Cohen and Witteborn 1985),
with the possible addition of some line-of-sight extinction
(Hanner et al. 1998).
We prefer to follow a different approach,
and compare the observed features to the predictions of specific models
of the circumstellar environment of TTS.
At present, the most interesting suggestion
on the origin on the silicate emission
is that of Calvet et al.~ (1992) and Chiang \& Goldreich (1997; CG97),
who point out that the feature could form in the  optically thin
surface layers of a circumstellar disk,
heated well above the temperature of the disk midplane
by the stellar radiation.
An alternative possibility has been suggested for the binary star
GW Ori, namely that the silicate emission originates in a disk ``gap",
cleared by the dynamical perturbation of a companion star. A small
amount of residual dust in the gap is heated by the stellar radiation
to temperatures of a few hundred degrees and causes the observed
emission (Mathieu et al. 1991).
This model predicts very little continuum
emission in the 10 \um\ region, as observed in GW Ori, where there
is clear evidence of a mid-infrared dip in the SED. This does not
seem to be the case of the TTS in our sample (Robberto et al. 1999),
and we will not discuss it further.

\subsection {Model Calculations}

We will follow here the very elegant and simple
treatment of CG97 to predict the 10 \um\ feature intensity
of each individual star, and we refer the reader to their paper for
a more detailed description of the model.
In CG97 models,
the stellar radiation penetrates the disk outer layers  to an
optical depth $\tau^\parallel \sim 1$ along the disk plane.
$\tau^\parallel$ is measured at the wavelength where
the stellar flux $F_\nu$ peaks (roughly at frequency $\nu_\star \sim$  2.1-3 $\times 10^{14}$ hz
(1--1.4 $\mu$m) for the stars in our sample).
In the direction
orthogonal to the disk surface, the optical depth at $\nu_\star$ is :

\begin{equation}
\tau^\perp = \tau^\parallel\> \alpha  \sim \alpha
\end{equation}

\noindent where
$\alpha$ is the grazing angle. If the disk is in hydrostatic equilibrium
in the vertical direction, $\alpha$ can be written as ($R\gg R_\star$):
\begin{equation}
\alpha \sim {{8}\over{7}} \Big({{T_\star}\over{T_c}}\Big)^{4/7}
   \Big({{R}\over{R_\star}}\Big)^{2/7}
\end{equation}
\noindent where R is the distance from the star and $T_c$ measures
the gravitational potential at the stellar surface:
\begin{equation}
T_c = {{GM_\star\,\mu}\over{KR_\star}}.
\end{equation}
\noindent K is the Stefan--Boltzman constant and $\mu$ is the
mean molecular weight.
This  superheated layer is optically thin in the orthogonal direction at
all wavelengths. Its emission can be easily computed as:
\begin{equation}
F_\nu^a = {1\over{D^2}} \int_{R_{in}} ^{R_{out}} {2\pi\,R\,dR\>\> B_\nu(T)
\>\>\tau_\nu^\perp }
\end{equation}
\begin{equation}
F_\nu^a = {{8}\over{7}}\,
{{2\pi R_\star^2}\over{D^2}}\> \Big({{T_\star}\over{T_c}}\Big)^{4/7}
\Big({{\sigma_\nu}\over{\sigma_\star}}\Big) \>\>\int _{R_{in}/R_\star}
^{R_{out}/R_\star} {B_\nu(T) \,{\rm x}^{9/7}\, d{\rm x}}
\end{equation}
where $D$ is the distance of the star, $B_\nu$ the Planck function,
$R_{in}$ and $R_{out}$ the inner and outer radius of the dusty disk and
$\tau_\nu^\perp = \alpha \times {{\sigma_\nu}\over{\sigma_\star}}$, with
$\sigma_\nu$ being the dust absorption cross section at $\nu$ and $\sigma_\star$
being the dust cross section at $\nu_\star$.
The dependence of $F_\nu$ on the disk viewing angle is weak
unless the disk is seen practically edge-on and  occults the emission
coming from the inner region of the disk itself  (Chiang and Goldreich 1999).

The dust temperature $T$ of grains in the optically thin
atmosphere depends on the stellar radiation
field and on the grain emissivity. For the cases we are considering,
we have checked using the radiation transfer code kindly provided
to us by E. Kr\"ugel that it can be
expressed as a power-law function of the
distance from the star $R$:
\begin{equation}
T \sim T_0 \Big ({{R}\over{R_\star}}\Big)^{-q}.
\end{equation}
\noindent  with
$q\sim 0.47$, $T_0\sim 0.75 T_\star$.
For the stars in our sample, $R_{in}$,
the distance from the star where the grains  sublimate ($T\sim 1500$ K),
varies between $\sim 4$ and $\sim 7 R_\star$, depending on the stellar
temperature. We have adopted $R_{out}=100$ AU; the results are not
sensitive to the exact value of these radii.

For each star, we have adopted the values of \Lstar, \Rstar, \Tstar
and \Mstar\ given in Table~1.
Once the stellar parameters are fixed, $F_\nu^a$
depends only on the dust cross section.
To clarify the following discussion, we re-write the ratio
$\sigma_\nu/\sigma_\star$ as the product
($\sigma_\nu/\sigma_{10}$).($\sigma_{10}/\sigma_\star$),
where $\sigma_{10}$ is the opacity at
the peak of the silicate feature.
The first term $\sigma_\nu/\sigma_{10}$ determines the shape
of the feature and depends only on the specific properties
of the silicates that dominate the feature
(size, mineralogy, and allotropic form).
The second term $\sigma_{10}/\sigma_\star$ determines the feature
intensity and is essentially the ``efficiency'' of converting stellar
photons into silicate emission.  This depends not only on the dust properties
enumerated above, but also on attributes which determine the relative dust
opacities from the visible/near--IR ($\sigma_\star$) to that near the feature
itself in the mid--IR ($\sigma_{10}$); for example, the relative contribution of carbonaceous
and silicate grains to $\sigma_\star$.
After a preliminary exploration of a variety of silicates (see \S 4.2),
we have adopted for $\sigma_\nu/\sigma_{10}$
the cross section of 1.2 \um\ radius pyroxene (Mg(0.5)Fe(0.43)Ca(0.03)Al(0.04)SiO$_3$;
Jaeger et al. 1994)
between 5.5 and 15 \um, but kept  as a free parameter
$\epsilon=\sigma_{10}/\sigma_\star$.
Note that
$\epsilon$ is the only free parameter of these models.

\begin{figure}
\insertplot{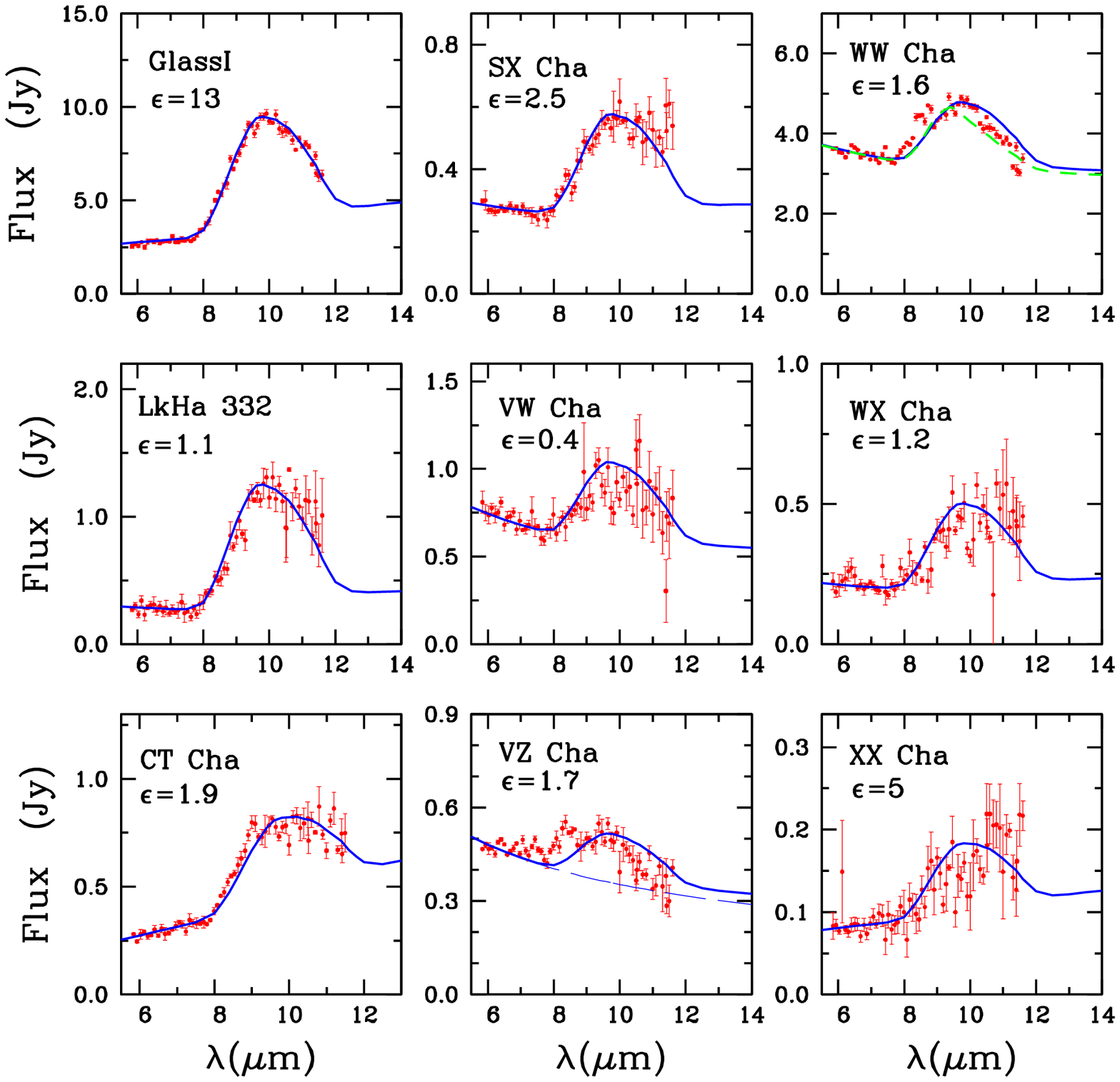}{8.0}{10.}{-0.75}{1.0}{0.8}{0}
\caption
{Model fits to the observed spectra. In each panel, the dots show the
observed points.
The solid line shows the predicted emission of the superheated
disk atmosphere, computed following Chiang \& Goldreich (1997) for
pyroxene grains of 1.2 \um\ radius, overimposed on the
continuum emission
shown by the dashed lines in Fig.~\ref{spectra_1}, \ref{spectra_2},
and \ref{spectra_3}.
In the \WW\ panel, the dashed line shows the model predictions for
small pyroxene grains (0.1 \um\ radius); in the \VZ\ panel,
the long-dashed line is the assumed continuum emission.
The  value of
$\epsilon =\sigma_{10}/\sigma_\star$ which gives the best
approximate fit to the observations is given in each panel.
}
\label{models}
\end{figure}

The results are shown in Fig.~\ref{models}.
In each panel, the dots show the observed spectrum.
The solid curve is the  model prediction, computed by adding to the
emission of the disk atmosphere the power-law continuum emission
shown by the dashed lines in Fig.~\ref{spectra_1}, \ref{spectra_2}, and
\ref{spectra_3}.
This characterization of the continuum emission is somewhat uncertain.
CG97 compute the emission of the optically thick
disk midplane in addition to that of the outer optically thin layers.
One could, in principle, use their model to obtain
values of the continuum emission consistent
with the structure of the disk atmosphere required to reproduce the
observed silicate emission. However, contrary to the atmospheric
emission in the 10 \um\ region, the contribution of the disk midplane to the
continuum depends on the inclination of the disk with respect to the
line of sight. Moreover,  at wavelengths $\lambda < 10 \ \mu$m
the disk midplane emission is more sensitive to viscous heating
and the inner disk radius (e.g. Meyer, Calvet, \& Hillenbrand, 1997).
Therefore, we defer the discussion of the
continuum emission (and of its consistency with the
results discussed in this paper) to a forthcoming
contribution (Robberto et al. 2000)
where we will make use of the broad-band fluxes measured by ISO over the
wavelength range 3.4 -- 200 \um\ (see Robberto et al. 1999 for
a presentation of some preliminary results).

The  value of $\epsilon=\sigma_{10}/\sigma_\star$ which provides the best
approximate fit to the observations  given the assumptions outlined
above is given
in each panel. Note that for each
star the model-predicted intensity of the feature (continuum subtracted)
is simply proportional to $\epsilon$.

\subsection {Mineralogy}

The assumption of the cross-section of 1.2 \um\ pyroxene grains
for $\sigma_\nu/s_{10}$ in the model calculations gives a
reasonable fit to the observed feature shapes. However, it is
by no means unique.
In our exploration of
the effects of dust mineralogy on the shape of the feature, we have
used as a reference the \Gl\ spectrum,
because of its high signal--to--noise;  the results would not be different for
the other stars.

\begin{figure}
\insertplot{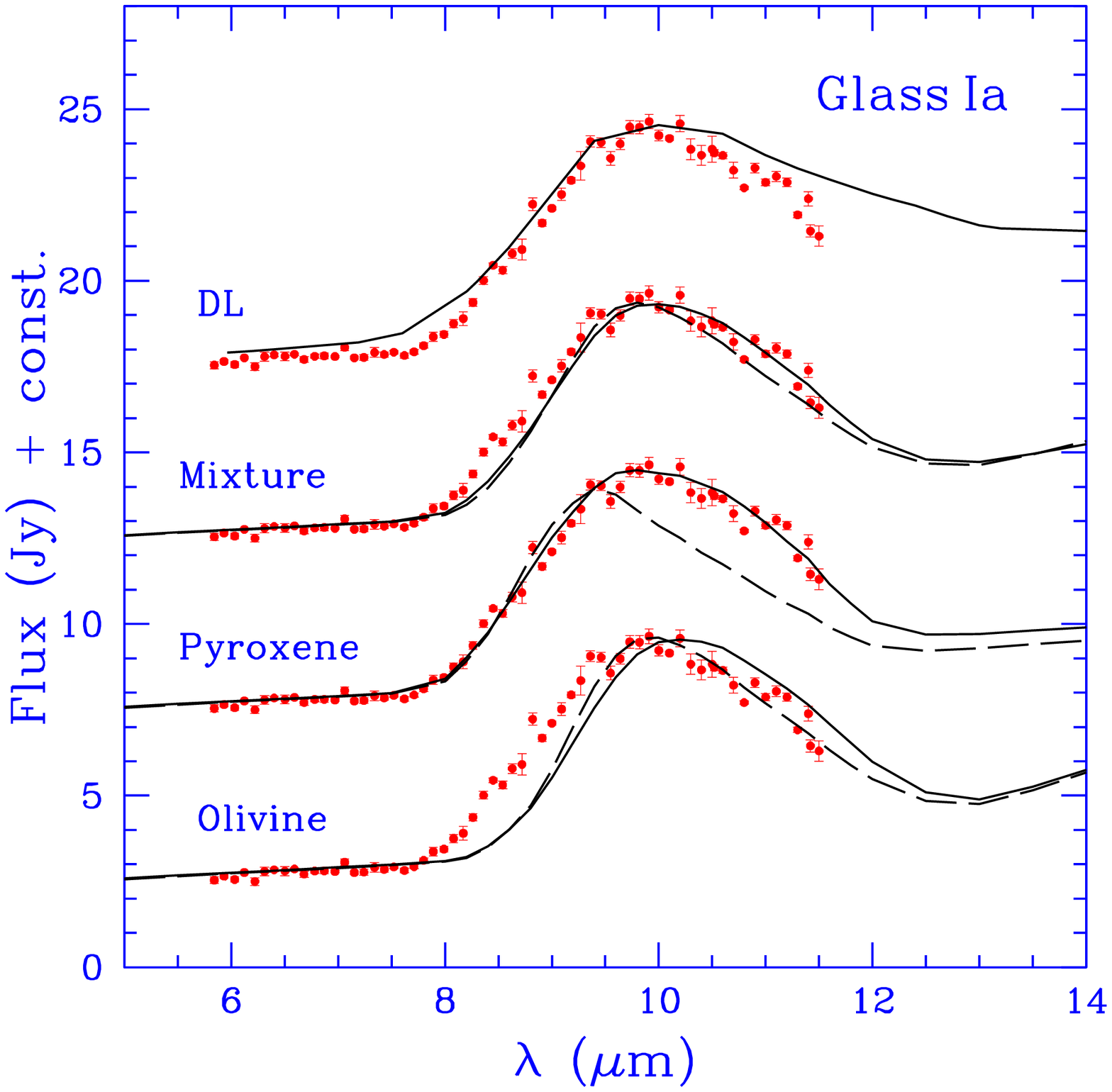}{8.0}{10.}{-0.75}{1.0}{0.8}{0}
\caption
{Model predictions for \Gl\ using different silicates.  From top
to bottom, we have:
DL astronomical silicates;
mixture of 0.5:0.5 olivine and pyroxene of 1 \um\ radius
(solid line) and 0.3:0.7 of 0.1 \um\ radius (dashed line);
pyroxene grains of radius 0.1 \um\ (dashed line),
1.2 \um\ (solid line):
olivine grains of radius
0.1 \um\ (dashed line) and 1 \um\ (solid line).
The model-predicted profiles have been
scaled to reproduce roughly the observed peak intensity.
The values of  $\epsilon$ required to fit the feature are all in the
interval 12--13.5. They can be compared
to the values derived from the cross sections
of these  various
minerals, which are   10 and 21 for pyroxene of 1.2 and 0.1 \um\ radius,
0.6 and 1.2 for 1 and 0.1 \um\ olivine,
2.2 for the 0.5:0.5 mixture of 1 \um\ grains, and 0.84 for the 0.3:0.7 mixture
of 0.1 \um\ grains.
The observations are shown
by dots.  }
\label{mineralogy}
\end{figure}

The model predictions for
\Gl\ and a variety of
amorphous silicates are shown in Fig.~\ref{mineralogy}.
We show results for
olivine of radius 0.1 and 1 \um\ (MgFeSiO$_4$; Dorschner et al. 1995);
for pyroxene grains of radius 0.1 and 1.2 \um\
((Mg(0.5)Fe(0.43)Ca(0.03)Al(0.04)Si O$_3$; Jaeger et al. 1994),
 and for mixtures of   olivine and pyroxene of 1 \um\ and 0.1 \um\ radius.
Olivine or small pyroxene
grains cannot  reproduce the feature shape;
a good fit is provided by rather large ($\sim 1$ \um)
 pyroxene grains and by
mixtures of olivine and pyroxene of  radii $\simless 1$ \um.
Much larger grains can be excluded, since they  give an emission
feature peaked at too long wavelengths.
The top panel of Fig.~\ref{mineralogy}
shows the \Gl\ model predictions for Draine and Lee (1984; DL)
astronomical silicates. The optical constants of such
 silicates were constrained
to reproduce the shape of the 10 \um\ feature observed by Forrest et al. (1979)
in the direction of the Trapezium stars for a MRN size distribution of grains,
i.e., for the case when the opacity is dominated by small grains (Mathis et al.,
 1977).  We find good agreement between the observations in the direction of
\Gl\ and DL astronomical silicates; the only significant
 discrepancy occurs at long wavelengths,
where the DL feature is  broader.
The Trapezium feature is significantly broader than that observed
in the diffuse ISM (see Draine, 1989).
The comparison with TTS seems to indicate
that silicates in the Trapezium region
are similar to those in TTS disks, and are
consistent with a mixture of amorphous olivines and pyroxenes.

Before discussing further our observations in the context of CG97 models,
let us point out that we may have some
evidence of two additional emission components at 8.5 \um\ and 11.3 \um.
Excess emission at long wavelengths
with respect to model predictions is seen in almost all stars.
A feature at 11.3 \um\
is typical of crystalline silicates, and it has been
detected in SWS spectra of some \pms\ stars
of intermediate mass (Malfait et al. 1999a,b). However, our observational
errors are larger at longer wavelengths, and the only clear case
where we see a secondary peak at $\lambda\simgreat$10 \um\ is
\Gl.  Also, the long-wavelength excess can be an artifact of our
crude fitting of the continuum between 5.8--8 $\mu$m.  This
can be checked by combining our PHOT-S data with observations
at longer wavelengths.  The spectrum of \Gl\ also shows that
a second feature may be present with peak at about 8.5 \um.
This feature is seen in SWS spectra of
some OH/IR and B[e] stars  (Waters, personal communication).
A possible attribution to SiO$_2$ is being investigated.
However, this feature sits near the 8.3 $\mu$m region
noted above for which the calibration is somewhat uncertain
and so should be treated with caution.
The absence of features at 6.2 and 7.6 \um\ rules out the
possibility that the excess emission at $\sim 8.5$ and 11.3 \um\
could be due to PAHs.

\subsection {Discussion}

The CG97 models are remarkably successful in accounting for the
observed emission.  For 6 stars,
$\epsilon$  is in the range 1--2.5.
We cannot determine if the spread of values from star to star within
this interval reflects  different grain properties or
uncertainties on the stellar parameters and inclination.
However, we can compare the values derived from our model with
those expected based on absorption cross--sections calculated
for various types of grains.
Amorphous olivine has $\epsilon \sim 1$ over a large range of sizes,
while pyroxene grains generally require larger values of $\epsilon$
($\sim 10$ for micron-size grains and $\sim 25$ or larger for smaller grains).
The olivine-pyroxene mixtures that reproduce the observed feature
shapes have $\epsilon \sim 1-2$.
Predictions for conglomerates and/or large
particles give values of $\epsilon$ in the range 1--1.5.
Henning and Stognienko (1996), for example,
give $\epsilon\sim 0.9-1.2$ for conglomerates made of pyroxene, iron,
and olivine. Pollack et al. (1994) compact grains, which include
a mixture of silicates, iron, organic materials and water ice,
have $\epsilon \sim 1-1.4$ (see Henning and Stognienko 1996).
In all these cases, the predicted shape of the feature should be
similar to the observed one.
It is also possible to reproduce the required $\epsilon$
value with  pyroxene grains,
if there is a contribution to the 1 \um\ opacity from carbonaceous grains.
However this contribution must be significantly smaller than in the interstellar medium; the MRN model of interstellar extinction  of Draine and Lee (1984) has
a 1 \um\ opacity dominated by graphite and
$\epsilon\sim 0.2-0.3$, depending on the exact value of \Tstar.
In summary,  amorphous olivine and pyroxene mixtures can roughly reproduce both
the intensity and the shape of the observed features for most of the TTS
in our sample. Large pure pyroxene grains cannot be excluded, as
long as some additional opacity at visual and near-infrared wavelengths
is provided by grains other than silicates.

In two cases we find much higher values of $\epsilon$, i.e.,
silicate emission much stronger than predicted by the CG97 models
with $\epsilon \sim 1-2.5$.
One (\XX) is a very faint object and
has a poorly measured spectrum, but the other
(\Gl)
is the strongest and best measured object in our sample.
\Gl\ has the exceptionally large  ratio
L$_{sil}$/\Lstar\ $\sim 10$\%, which requires for the stellar
parameters given in Table 1 $\epsilon\sim 13$.
Pyroxene grains of radius 1--1.2 \um\ (which  fit well the shape of
the feature) have
$\epsilon \sim 10$. It is tempting to speculate that very strong
silicate features may occur when the dust opacity
is dominated by pyroxene grains over the whole range of wavelengths
$\sim 1-10$ \um.
However, we should point out that
\Gl\ is a rather complex object.
It is a binary system, with an infrared companion (IRC; component b)
at a separation of about
2.7$"$ (Chelli et al. 1988).
 Feigelson and Kriss (1989) report that this
faint companion is an emission--line star of intermediate spectral type
(G3--G7).  Although component b is very faint in the visible, it dominates
the emission at wavelengths longer than 2.0 \um\ (Chelli et al. 1988).
There are a handful of such objects known
in T Tauri systems, where the IRC is responsible for the bulk of the IR
emission and is thought to be in an earlier evolutionary state.
Recent, unpublished 10 \um\ broadband images (Robberto et al. and
Stanke and Zinnecker, private
communications) confirm that the 10 \um\ emission comes from a single source,
very likely Glass Ib.
We have computed the silicate emission of Glass Ib, assuming
a {\it stellar} luminosity (neglecting the contribution of active accretion)
of 3.2 \Lsun, an effective temperature of 5400 K and a mass of 2 \Msun
(e.g. Koresko et al. 1997); the value
of $\epsilon$ we derive is $\sim 12$, similar to that obtained under the
assumption that Glass Ia is powering the feature.

A comparison of silicate spectra obtained for \Gl\ with observations of
other T Tauri+IRC systems
sheds no light on the matter.  Spatially unresolved observations of
T Tauri N \& S indicate a nearly featureless 10 \um\ continuum
(Hanner et al. 1998).  However, spatially resolved 10 $\mu$m observations
show that silicate emission is associated with the visible star,
while the IR companion has silicate in absorption (Herbst et al. 1997).
SX Cha, the other star in our sample with an IRC, has a rather normal silicate emission feature.
It is possible that in \Gl\ the silicate emission forms not in a disk, but in
a more spherically distributed warm dust component.
We cannot solve this puzzle with
the data presently available.  Any model which adequately explains the silicate
emission feature in this system, must also explain its
variability (cf. G\"urtler et
al. 1999).   For the moment, the best explanation is that in
Glass Ia or in Glass Ib the
dust has an anomalous composition, perhaps
including a small contribution from
crystalline silicates.

Only one star in our sample (\VW) has a very small ratio
L$_{sil}$/\Lstar and a low value of
$\epsilon$ (and a rather noisy spectrum).
There are a number of effects that, in principle, can result
in a silicate feature somewhat weaker than predicted by CG97 models.
Firstly, the disk flaring may be smaller than assumed in
CG97. This may happen, for example, if
the dust is not well mixed
with the gas, or if  the temperature gradient in the vertical direction
affects the density profile, which CG97
compute from the condition of hydrostatic equilibrium
of an isothermal gas (see discussion in Kenyon and Hartmann 1987).
Second, in cases where the disk is seen nearly edge--on,
one does expect to observe silicate features in absorption
as observed in a few cases by  Cohen and Witteborn (1985)
and as modelled by Chiang and Goldreich (1999).
A third important point is that
heating of the disk by viscous dissipation of accretion
energy may invert the vertical temperature gradient, so that the
temperature is higher in the disk midplane than in the outer layer.
In this extreme case, the feature should appear in absorption
(Calvet et al. 1992). In reality, this is probably rarely the case,
since about half of the accretion luminosity is emitted as UV radiation
near the stellar surface and may contribute to the radiative heating
of the disk outer layer (D'Alessio et al. 1998). In fact,
FU Ori, where the accretion luminosity is much larger
than the photospheric one, has very weak silicate emission
(Hanner et al. 1998).
A study of the SED of \VW\ over the whole range from
UV to millimeter wavelengths, which can shed light on its properties,
is deferred to a forthcoming paper (Robberto et al. 2000).

To summarize, the comparison with CG97 models
provides a strong indication that;
i) most TTS  disks must be flared
approximately as predicted by hydrostatic equilibrium models;
ii) accretion,  when present,
does not dominate the disk heating, which is mostly due
to stellar radiation at radii $\gg 0.1$ AU;
and iii) the viewing geometry for the stars in our sample is
not perfectly edge--on.

\section {SUMMARY AND CONCLUSIONS}

We have presented in this paper PHOT-S spectra of nine
classical TTS in Chamaeleon. The spectra cover the
wavelength ranges 2.5--4.9 and 5.9--11.7 \um.
The only prominent feature in the spectra is the silicate emission
feature at about 10 \um, which is seen in all the observed stars.

\vfill\eject

We discuss the possibility that the feature forms in the outer,
optically thin layers of a circumstellar disk. If the disk heating is
dominated by the stellar radiation, this disk atmosphere is hotter
than the disk midplane, and emission features may be seen.
We have followed the description of the disk atmosphere  developed by
Chiang and Goldreich (1997), and computed the predicted
silicate feature for each star in our sample.
We find  good agreement between the model predictions
and the observations for most stars.

In these  models,
the shape of the feature depends on the properties
(size, chemical composition and allotropic form)
of the silicates that dominate the
10 \um\ opacity, while its intensity
depends on
the ratio ($\epsilon$) of the opacity
 at the peak of the feature to the opacity at the wavelength where
the stellar radiation is maximum.
These are the only free parameters of the models, once the stellar
properties (luminosity, effective temperature and mass) are known.
For six stars out of nine a good fit between models and observations is obtained
with values of $\epsilon\sim 1-2.5$, much larger
than in the diffuse ISM, where $\epsilon \sim 0.2-0.4$.
This, and the analysis of the shape of the features,
suggests that silicates in the TTS disks are likely to be a mixture of
amorphous olivine and pyroxene grains with radii $\simless 1$ \um,
with little contribution to the visual and near-infrared opacity from
other dust components. Alternatively, our observations can be reproduced
by large ($\sim 1$ \um) pyroxene grains, with a significant contribution
to the short-wavelength opacity from grains other than silicates.

The identification of disk atmospheres as the origin of silicate emission
is certainly not a unique explanation. Full disk models, capable of reproducing
both the silicate spectrum and the observed continuum over a large
range of wavelengths need to be computed. Disk parameters,
such as inclination, accretion luminosity, etc., have to
be consistent with what we know about each individual star.
Broad-band PHOT observations of
the same TTS discussed here  over the wavelength range 3.4-200 \um\
have been obtained and are currently being investigated (Robberto et al. 2000).
However, the results presented here  give already  some
interesting information on the disk properties. As discussed in \S 4,
we have assumed that the disk is  in hydrostatic equilibrium in
the vertical direction (flared). Atmospheres of flat disks are much
less extended, and will reproduce the observed silicate emission
only for unreasonably large values of $\epsilon$ ($\simgreat 100$).
Secondly, strong heating due to accretion in the disk midplane
will reduce or suppress the inverse temperature gradient
in the vertical direction and reduce the strength of the silicate emission.
In the stars we observed, accretion is likely to contribute only
marginally to the disk heating at radii $>$ 0.1 AU.

\acknowledgements

We gratefully acknowledge Bruce Draine, Thomas Henning,
Malcolm Walmsley, and Endrik Kr\"ugel
for helpful discussions, the referee for a helpful and
informative review of this work, and Greet Decin, Bart Vandenbussche, J\"orgen
G\"urtler, and Rens Waters for sharing information regarding PHOT--S data.
Special thanks to Jose Acosta-Pulida  as well as
Peter Abraham for assistance with the data reduction
and calibration procedures.  This work was partly supported by ASI grant
ARS-98-116 to the Osservatorio di Arcetri.  Support for MRM was provided
by NASA through Hubble Fellowship Grant \# HF--01098.01--97A awarded by the
Space Telescope Science Institute, which is operated by the Association of
Universities for Research in Astronomy, Inc., for NASA under
contract NAS 5--26555.


\vfill\eject

\clearpage
\vfill\eject


\begin{table}
\begin{small}
\caption{Stellar Parameters}
\begin{tabular}{llccccccc} \hline \hline
\multicolumn{1}{c}{Name} &
\multicolumn{1}{c}{SP$^1$} &
\multicolumn{1}{c}{T$_\star$ (K)} &
\multicolumn{1}{c}{EW(H $\alpha)^1$ (\AA)} &
\multicolumn{1}{c}{Compan.$^2$} &
\multicolumn{1}{c}{L$_\star^3$ (L$_\odot$) } &
\multicolumn{1}{c}{R$_\star^3$ (R$_\odot$) } &
\multicolumn{1}{c}{M$_\star^4$ (M$_\odot$) } &
\multicolumn{1}{c}{Age$^4$ (Myr) } \\\hline\hline
CT~Cha  & K7   & 4000 & 49.2  & NC$^a$           & 0.7 & 1.8  & 0.5 & 0.9 \\
Glass~Ia& K4   & 4600 & 4.0   & 2.8''/0.32$^{b}$ & 1.6 & 2.1  & 0.9 & 1.0 \\
LkH$\alpha$~332-20& K2& 4900  & 43.6 & NC$^a$    & 3.3 & 2.6  & 1.2 & 1.0 \\
SX~Cha  & M0.5 & 3700 & 26.7  & 2.1''/4.7$^c$    & 0.5 & 1.7  & 0.4 & 0.8 \\
VW~Cha  & K5   & 4350 & 146.9 & 0.7''/4.5$^a$    & 2.9 & 3.1  & 0.6 & 0.3 \\
VZ~Cha  & K6   & 4200 & 71.4  & NC$^a$           & 0.5 & 1.3  & 0.7 & 3.0 \\
WX~Cha  & K7-M0& 3900 & 65.5  & 0.8''/10$^a$     & 0.8 & 2.0  & 0.4 & 0.7 \\
WW~Cha  & K5   & 4350 & 67.4  & NC$^a$           & 2.7 & 3.0  & 0.6 & 0.3 \\
XX~Cha  & M1   & 3650 & 133.5 & --               & 0.1 & 0.85 & 0.5 & 7.0 \\\hline
\multicolumn{9}{l}{\footnotesize $^1$ Spectral information taken from
Gauvin \& Strom (1992). } \\
\multicolumn{9}{l}{\footnotesize $^2$
Presence or absence of companions ($d < 14$'') taken from a)
Ghez et al. (1997); b) Chelli et al. (1988); or } \\
\multicolumn{9}{l}{\footnotesize
c) Robberto et al. (2000). For known doubles, the separation is given in
arcseconds followed by the } \\
\multicolumn{9}{l}{\footnotesize
 2.2 $\mu$m flux ratio [F$_{2.2}$
(optical primary)/F$_{2.2}$(IR companion)].  } \\
\multicolumn{9}{l}{\footnotesize $^3$
Stellar luminosity is calculated from bolometric corrections applied
to dereddened I--band magnitudes } \\
\multicolumn{9}{l}{\footnotesize following Meyer (1996) and the stellar
radius is derived from the Stefan--Boltzman equation. } \\
\multicolumn{9}{l}{\footnotesize $^4$
Masses and ages are derived from comparison of stellar effective
temperatures and luminosities } \\
\multicolumn{9}{l}{\footnotesize
with the pre--main sequence evolutionary tracks of
D'Antona \& Mazzitelli (1994) using CM } \\
\multicolumn{9}{l}{\footnotesize
convection and Alexander opacities.} \\
\label{properties}
\end{tabular}
\end{small}
\end{table}

\begin{table}
\begin{small}
\centering
\caption{Properties of the Silicate Feature}
\begin{tabular}{lccc} \hline \hline
Name  &L$_{sil}$& L$_{sil}$/L$_\star$& $F_{peak}/F_{cont}$\\
       & ($10^{-2}$L$_\odot$)&  (\%)  &  \\ \hline
CT~Cha & 1& 1.4& 2.2\\
Glass~Ia& 17 & 10& 2.6\\
LkH$\alpha$~332-20& 3&2& 3.7\\
SX~Cha& 1& 0.8& 1.4\\
VW~Cha& 0.7& 0.2& 0.4\\
VZ~Cha& 0.2& 0.5& 0.2\\
WX~Cha& 0.8& 1& 1.7\\
WW~Cha& 5& 1.8 & 0.5\\
XX~Cha& 0.2& 2& 1.2\\
\end{tabular}
\end{small}
\end{table}

\clearpage
\vfill\eject


\end{document}